\documentclass[fleqn,usenatbib]{mnras}

\usepackage{newtxtext,newtxmath}
\usepackage[T1]{fontenc}

\DeclareRobustCommand{\VAN}[3]{#2}
\let\VANthebibliography\thebibliography
\def\thebibliography{\DeclareRobustCommand{\VAN}[3]{##3}\VANthebibliography}

\usepackage{graphicx}	% Including figure files
\usepackage{amsmath}	% Advanced maths commands
\title[Energy equipartition]{Effects of massive central objects on the degree of energy equipartition of globular clusters.}

\author[Aros \& Vesperini]{
Francisco I. Aros$^{1}$\thanks{E-mail: faros@iu.edu},  Enrico Vesperini$^{1}$
\\
$^{1}$Department of Astronomy, Indiana University, Bloomington, Swain West, 727 E. 3rd Street, IN 47405, USA\\
}

\date{Accepted XXX. Received YYY; in original form ZZZ}

\pubyear{2023}

\begin{document}
\label{firstpage}
\pagerange{\pageref{firstpage}--\pageref{lastpage}}
\maketitle

\begin{abstract}
We present an analysis of the degree of energy equipartition in a sample of 101 Monte Carlo numerical simulations of globular clusters (GCs) hosting either a system of stellar-mass black holes (BHS), an intermediate-mass black hole (IMBH) or neither of them. For the first time, we systematically explore the signatures that the presence of BHS or IMBHs produces on the degree of energy equipartition and if these signatures could be found in current observations. We show that a BHS can halt the evolution towards energy equipartition in the cluster centre. We also show that this effect grows stronger with the number of stellar-mass black holes in the GC. The signatures introduced by IMBHs depend on how dominant their masses are to the GCs and for how long the IMBH has co-evolved with its host GCs. IMBHs with a mass fraction below $2\%$ of the cluster mass produce a similar dynamical effect to BHS, halting the energy equipartition evolution. IMBHs with a mass fraction larger than $2\%$ can produce an inversion of the observed mass-dependency of the velocity dispersion, where the velocity dispersion grows with mass. We compare our results with observations of Galactic GCs and show that the observed range of the degree of energy equipartition in real clusters is consistent with that found in our analysis. In particular, we show that some Galactic GCs fall within the anomalous behaviour expected for systems hosting a BHS or an IMBH and are promising candidates for further dynamical analysis.
\end{abstract}


%%%%%%%%%%%%%%%%%%%%%%%%%%%%%%%%%%%%%%%%%%%%%%%%%%

%%%%%%%%%%%%%%%%% BODY OF PAPER %%%%%%%%%%%%%%%%%%

\section{Introduction}

Dense and compact stellar systems such as globular clusters (GC) evolve dynamically through stellar encounters in a process known as two-body relaxation. Two of the main main manifestations of the effects of two-body relaxation in GCs are mass segregation \citep{spitzer_1975,spitzer_1987,heggie_hut_2003} and the evolution towards energy equipartition \citep{spitzer_1969,spitzer_1987,heggie_hut_2003}. As a result of mass segregation, GCs develop a radial dependence of the stellar mass, with the mean mass of stars in the cluster increasing towards its centre. This is due to high-mass stars sinking towards the cluster centre as they transfer kinetic energy to low-mass stars that move outward in the cluster. 

The interchange of kinetic energy between stars of different masses makes the GC evolve towards energy equipartition; however, whether a star cluster may achieve full energy equipartition or not depends on the stellar mass function and, as shown by \citeauthor{trenti_2013} (\citeyear{trenti_2013}, see also \citeauthor{vishniac_1978} \citeyear{vishniac_1978}), GCs starting with a realistic initial mass function (IMF) do not achieve full energy equipartition in their lifetime. The degree of energy equipartition in a cluster is a local effect and is traced by the velocity dispersion of stars with different masses within the same region. If no equipartition is present, then the velocity dispersion of the stars is independent of their masses. Once low- and high-mass stars interchange kinetic energy, the velocity dispersion of low-mass stars will increase while the velocity dispersion of high-mass stars will decrease. This mass-dependent velocity dispersion traces the degree of energy equipartition. In the case of full equipartition, the velocity dispersion decreases with the mass following a log-slope of $-0.5$ \citep[see e.g.][]{spitzer_1969,trenti_2013,bianchini_2016}. At a global level, the degree of energy equipartition increase towards the cluster centre, following the higher stellar densities and shorter relaxation times.

Different kinematic properties of a GC, such as internal rotation or velocity anisotropy \citep[see e.g.][]{einsel_1999,hong_2013,breen_2017,pavlik_2021,pavlik_2022,livernois_2022,kamlah_2022}, or the presence of primordial binaries \citep[see e.g.][]{vesperini_1994,trenti_2007,chatterjee_2010,bhat_2023} can affect its dynamical evolution.

The presence of multiple stellar-mass black holes in a GC (black hole systems, BHS) or a single intermediate-mass black hole (IMBH) can also impact the GC evolution. They provide additional kinetic energy to the cluster, delaying its dynamical evolution and slowing the effects of mass segregation and energy equipartition \citep[see][]{baumgardt_2004,mackey_2008,gill_2008}. Previous studies show that by measuring the mass segregation in GCs is possible to infer the presence of a BHS \citep{alessandrini_2016,weatherford_2018,weatherford_2020} or IMBH \citep{pasquato_2009,pasquato_2016,de_vita_2019}, both from the simulation and observational side. This requires faint enough photometric samples to characterise the radial distribution of stars with different masses along the main sequence. Measuring the degree of energy equipartition in a GC, on the other hand, requires kinematic data for a large enough sample of different stellar masses and is particularly challenging at fainter magnitudes. With the continuous improvement of kinematic data from space telescopes, particularly HST, studies in the last decade have begun to be able to measure the degree of energy equipartition \citep[see][]{heyl_2017,libralato_2018,libralato_2019}. More recently, \cite{watkins_2022} measured the degree of energy equipartition in a sample of $8$ Galactic GCs, not only for the case considering all stars within the field of view (generally about the half-light radius) but also producing for the first time radial profiles for the degree of energy equipartition in the clusters. While most clusters show an increasing degree of energy equipartition towards the cluster centre, a couple of GCs show a flat profile or even cases when the degree of energy equipartition is lower at the innermost region (see their Figure 14, particularly for NGC 5139, NGC 6266 and NGC 6752). Motivated by these results, we turned to numerical simulations of GCs to explore which dynamical effects can produce this variety of radial profiles and what we can learn about the dynamical evolution of the GC from these kinds of observations. 

In this work, we analyse the dynamical effects of a BHS or an IMBH in the degree of energy equipartition for a sample of simulated GCs, taking into account observational constraints and the presence of kinematic errors. We follow the approach presented by \cite{watkins_2022}. to obtain the degree of energy equipartition from discrete kinematic data, also aiming to put the measurements from simulations and observations in the same comparable framework. 

In Section \ref{sec:sim}, we introduce the sample of simulated globular clusters and the selection of stars used in our analysis. We introduce the models that describe the degree of energy equipartition and the fit to the discrete stellar kinematics from \cite{watkins_2022} in Section \ref{sec:models}. In Section \ref{sec:results}, we present our main results. We discuss the cases with a BHS and an IMBH further in Section \ref{sec:dyn_bh}. Finally, we summarise our work in Section \ref{sec:summary}.% and describe the next steps for the results presented here.

\section{Simulations}
\label{sec:sim}

For our analysis, we use a sample of 101 simulated GCs evolved to $12\,\text{Gyr}$ using the \texttt{MOCCA} code \citep[see ][]{hypki_2013,giersz_2013}. These simulations are part of the \textsc{MOCCA-SURVEY} Database I \citep{belloni_2016,askar_2017}, an extensive library of about $2000$ simulated GCs with different initial conditions \citep[see][]{askar_2017}. The \texttt{MOCCA} code follows the evolution of clusters including the effects of two-body relaxation, binary star interaction and a tidal truncation calculated assuming the cluster is moving on a circular orbit in a point mass potential. The \texttt{SSE} and \texttt{BSE} codes \citep{hurley_2000,hurley_2002} are used to model the effects of binary and stellar evolution; binary-single and binary-binary interactions are modeled with the the \texttt{FEWBODY} code \citep{fregeau_2004}. The models evolved with \texttt{MOCCA} are spherically symmetric and do not include any internal rotation.

In this paper we focus our attention on a subset of models that include GCs that have survived $12\,\text{Gyr}$ of evolution, with initial binary fractions of $5\%$ and $10\%$, and with masses that allow for a velocity dispersion larger than $4\,\text{km}\,\text{s}^{-1}$ at their 2D half-light radius ($R_h$) in the $t=12\,\text{Gyr}$ snapshot. Our sample, in particular, includes models that started with: $N=\{4\times10^5,7\times10^5,1.2\times10^6\}$ stars, a \cite{king_1966} model with central dimensional potential $W_0=\{3,6,9\}$, tidal to half-mass radius ratio of $r_t/r_{50\%}=\{25,50\}$ including filling models with $r_t/r_{50\%}\lesssim10$ , and metallicities of $Z=\{0.001,0.005,0.02\}$. All models follow a Kroupa IMF \citep{kroupa_2001} with masses between $0.08\,M_{\odot}$ and $100\,M_{\odot}$ \citep[see][ for a detailed description of the initial conditions considered in the \textsc{MOCCA-SURVEY} Database I]{askar_2017}. The models selected include clusters that at $12\,\text{Gyr}$ host a central IMBH or multiple stellar-mass black holes.

As described in \cite{giersz_2015}, some of the models in the \textsc{MOCCA-SURVEY} Database I are characterized by the presence of a central IMBH that may either form rapidly during the early cluster evolution ($t<100\,\text{Myr}$) or later when the system reaches high central densities during the cluster core-collapse phase \citep[see][for further details and discussion]{giersz_2015}. Regardless of the mechanism leading to the growth of an IMBH, the important point for our analysis is that we have models hosting an IMBH from the very early evolutionary phases of evolution and others in which the IMBH forms only later after a significant fraction of a cluster dynamical evolution. $41$ of $101$ models selected for this paper have an IMBH at $12\,\text{Gyr}$.

Models in the \textsc{MOCCA-SURVEY} Database I follow two prescriptions for the natal kicks velocities of stellar-mass black holes: (1) a Maxwellian distribution with $\sigma=265\,\text{km}\,\text{s}^{-1}$ \citep{hobbs_2005} or (2) a “fallback” prescription \citep{belczynski_2002} which allows for significant retention stellar-mass black holes within the first $20$ to $30\,\text{Myr}$ of the cluster’s evolution. In our sample, all clusters that host multiple stellar-mass black holes at $12\,\text{Gyr}$ follow the “fallback” prescription, while most clusters without multiple stellar-mass black holes follow the Maxwellian prescription. For further details on the stellar-mass black hole retention in the \textsc{MOCCA-SURVEY} Database I, see \cite{askar_2017,askar_2018}.

\subsection{Observed sample from simulations}
\label{sec:observables}

In our analysis, we study the degree of energy equipartition and its variation with the distance from the cluster's centre both using three-dimensional quantities (3D radial clustercentric distance and the three spherical components of the velocity) and two-dimensional projected quantities (projected radial distance $R$, the radial proper motion $v_{\text{pmR}}$ and tangential proper motion $v_{\text{pmT}}$). The analysis based on projected quantities is aimed at establishing a closer connection with observational studies. 

In each GC we select a subsample of only main sequence stars. To do so, we set a mass range between $0.2\,M_{\odot}$ and the stellar mass at the main sequence turn-off (MSTO), which corresponds to stellar masses between $0.8\sim1.0\,M_{\odot}$. The differences between the masses at the MSTO in our GC sample are primarily a consequence of the different metallicities of the clusters and, in a second order, the product of binary interactions such as mergers and mass transfer, which are stochastic effects. For the three metallicity values we find that the mean MSTO masses at $12\,\text{Gyr}$ are: $0.82\,M_{\odot}\,(z=0.001)$, $0.87\,M_{\odot}\,(z=0.005)$ and $0.94\,M_{\odot}\,(z=0.02)$. We only consider stars within a range of $0.02$ magnitudes in colour from the main sequence's mean colour; this selection excludes most binaries in the subsample. We do not include any restriction on the position of the observed stars, as we also aim to analyse the radial variation of the energy equipartition.

In order to account, at least in part, for the effects of observational errors, we added observational errors to the proper motion following an error distribution from available kinematic data of Galactic GCs. We used photometric and kinematic data of NGC 6752 from the HACKS catalogue\footnote{HACKS (Hubble Space Telescope Atlases of Cluster Kinematics) is available at \url{https://archive.stsci.edu/hlsp/hacks}.} \citep{libralato_2022} to get the median proper motion error at different magnitudes and construct a magnitude-dependent error function (see Figure \ref{fig:kin-err} in the Appendix \ref{app:errors}). The kinematic errors vary between $0.054\,\text{mas}\,\text{yr}^{-1}$ at the MSTO and $0.163\,\text{mas}\,\text{yr}^{-1}$ at seven magnitudes below the MSTO, which, at the distance of NGC 6752, corresponds to $1.1\,\text{km}\,\text{s}^{-1}$ and $3.2\,\text{km}\,\text{s}^{-1}$, respectively. The added errors translate to a range between $15\%$ to $35\%$ of the GCs velocity dispersions within the half-light radius. With this choice of observational errors, we assume that all simulated clusters in our sample are observed as if they were at the distance of NGC 6752 \citep[$d=4.125\,\text{kpc}$,][]{baumgardt_2021}.

\section{Measuring the Energy Equipartition}
\label{sec:models}

We study the degree of equipartition on each simulated GC using the two main descriptions used in the literature:
\begin{equation}
    \sigma(m) = \sigma_{0}\left(\frac{m}{m_{0}}\right)^{-\eta}\,,
    \label{eq:eta}
\end{equation}
where $\eta = 0$ indicates no equipartition and $\eta=0.5$ indicates full equipartition \citep[see e.g.][]{trenti_2013}. We use $m_{0} = 1M{\odot}$ so that $\sigma_{0}$ is the velocity dispersion of stars with masses of $1M_{\odot}$. 

The second commonly used description for the degree of equipartition \citep{bianchini_2016} is given by:
\begin{equation}
    \sigma(m) = 
    \begin{cases}
     \sigma_{0}\exp{(-0.5m/m_{\text{eq}})}  & \quad, m \leq m_{\text{eq}}\\
    \sigma_{\text{eq}}(m/m_{\text{eq}})^{-0.5}  & \quad, m > m_{\text{eq}},
    \end{cases}
    \label{eq:meq_original}
\end{equation}
where stars with masses lower than $m_{\text{eq}}$ (equipartition mass) follow a exponential form, and stars with masses larger than $m_{\text{eq}}$ are in full equipartition (following the same parameterization of Eq. \ref{eq:eta}). The normalization parameter $\sigma_{\text{eq}}$ corresponds to the velocity dispersion for stars with masses equal to the equipartition mass $m_{\text{eq}}$ and relates to the normalization parameter $\sigma_{0}$ through $\sigma_{\text{eq}} = \sigma_{0}\exp{(-0.5)}$. 

\begin{figure*}
    \centering
    \includegraphics[width=1.0\linewidth]{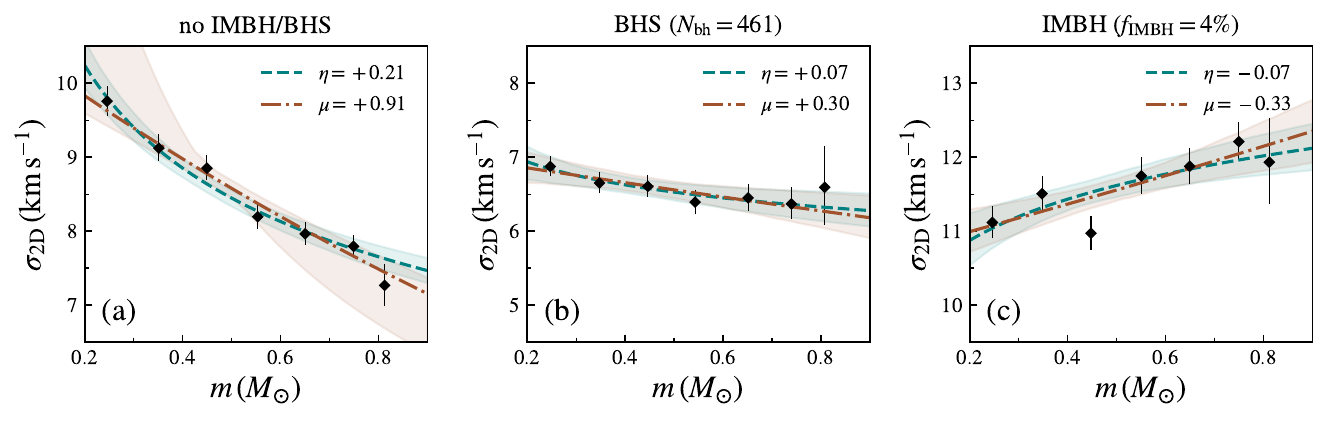}
    \caption{Velocity dispersion as function of stellar mass in three clusters for stars within the inner $10\%$ of their respective projected half-light radius. The three GCs have different central dark components: (a) no black holes,  (b) $461$ stellar-mass black holes and (c) an IMBH having $4\%$ of the cluster mass. In each case, we show the best-fit models for the degree of energy equipartition given the two representations used in this work: $\eta$ and $\mu$ (given Equations \protect\ref{eq:eta} and \protect\ref{eq:mu}, respectively). For (a) and (b), the velocity dispersion follows the expected behaviour with increasing values for lower masses. However, for (c), $\eta$ and $\mu$ have negative values resulting from the presence of an IMBH. The clusters in panels (a) and (b) have the same initial conditions, but different stellar-mass black hole retention prescription.}
    \label{fig:example_fit}
\end{figure*}

As we will discuss later, the presence of an IMBH can alter the slope of the mass-dependent velocity dispersion and lead to a velocity dispersion increasing with mass. The positive slope translates to a negative value of $\eta$ (see panel (c) in Figure \ref{fig:example_fit}). To characterise the positive slope, we first allow for negative $\eta$ values and use the same expression of Eq \ref{eq:eta}. On the other hand, we need to redefine the description of equipartition given by Eq. \ref{eq:meq_original} and allow for a “\textit{negative}” equipartition mass. In Eq. \ref{eq:meq_original}, the slope is steeper when $m_{\text{eq}}$ is positive and closer to zero, then while $m_{\text{eq}}$ increases, the slope becomes flatter until becoming zero at $m_{\text{eq}}\rightarrow\infty$. In the same way, when $m_{\text{eq}}$ goes from $-\infty$ to $0$, the slope becomes positive and steeper as $m_{\text{eq}}$ approaches $0$ from the left. To have a continuous representation of the change in the slope of Eq \ref{eq:meq_original}, we define $\mu = 1/m_{\text{eq}}$ and redefine Eq \ref{eq:meq_original} as:
\begin{equation}
    \sigma(m) = 
    \begin{cases}
    \sigma_{0}\exp{(-0.5m\mu)}  & \quad, \mu < 0\\
    \sigma_{0}\exp{(-0.5m\mu)}  & \quad, \mu \geq 0\,\,\text{and}\,\,m\mu < 1 \\
    \sigma_{\text{eq}}(m\mu)^{-0.5}  & \quad, \mu \geq 0\,\,\text{and}\,\, m\mu > 1.
    \end{cases}
    \label{eq:mu}
\end{equation}

We have that the slope of the degree of energy equipartition follow the parameterizations as:
\begin{equation}
\begin{array}{c|c|c}
  \text{Positive slope} & \text{Flat slope} & \text{Negative slope} \\
  \eta < 0    & \eta = 0 & \eta>0 \\
  0 > m_{\text{eq}} > -\infty & m_{\text{eq}} = -\infty,+\infty & 0 < m_{\text{eq}} < +\infty \\
  \mu < 0 & \mu = 0 & \mu>0,
\end{array}
\label{eq:new_range}
\end{equation}

For each GC, we estimate the degree of energy equipartition following Eq. \ref{eq:eta} and \ref{eq:mu} and the discrete likelihood approach implemented by \cite{watkins_2022}. In the case of the projected subsample, the likelihood function can be written as follows: 
\begin{equation}
\begin{split}
L(v_{\text{pmR}},&v_{\text{pmT}},m\mid\Theta) = \prod_{i}^{N}(2\pi(\sigma(m_i\mid\Theta)^2+\delta_{\text{pmR},i}^2))^{-1/2}\times\\
&(2\pi(\sigma(m_i\mid\Theta)^2+\delta_{\text{pmT},i}^2))^{-1/2}\times\\
&\exp\left(-\frac{1}{2}{\frac{v_{\text{pmR},i}^2}{\sigma(m_i\mid\Theta)^2+\delta_{\text{pmR},i}^2}-\frac{1}{2}\frac{v_{\text{pmT},i}^2}{\sigma(m_i\mid\Theta)^2+\delta_{\text{pmT},i}^2}}\right)\,,    
\end{split}
\label{eq:likelihood}
\end{equation}
where $m_i$ is the mass of the i-th star, $v_{\text{pmR},i}$ and $v_{\text{pmT},i}$ are the radial and tangential proper motions of the i-th star with observational errors given by $\delta_{\text{pmR},i}$ and $\delta_{\text{pmT},i}$, respectively. $\Theta$ encapsulates the parameters of the energy equipartition model $\sigma$ (Equations \ref{eq:eta} and \ref{eq:mu}). 

The likelihood function considers both proper motions together without including the effects of the velocity anisotropy. This is not an issue within the inner regions of the cluster, as they are consistent with being isotropic. However, for regions outside of the half-light radius, the velocity anisotropy becomes radial for a significant fraction of our GCs sample. We further discuss the implications of the velocity anisotropy in section \ref{sec:summary}. We utilise a similar likelihood function for the spherical velocity components as in Eq. \ref{eq:likelihood}, but with the three velocities and excluding velocity errors. 

Figure \ref{fig:example_fit} shows three GCs in our sample and their velocity dispersion dependence on mass, which traces the degree of energy equipartition in clusters. The “no IMBH/BHS” and “BHS” have the same initial conditions with the exception of the natal-kick prescription for stellar-mass black holes. The “IMBH” cluster has an IMBH with a mass fraction of $4\%$ of the cluster mass and is shown here since it provides a clear example of the trend between velocity dispersion and mass opposite to that expected from energy equipartition. We include stars in subsamples within $10\%$ of the half-light radius for each cluster, and use the likelihood function in Eq. \ref{eq:likelihood} to estimate the degree of equipartition. For both cases, we include the best fit as a dashed ($\eta$ model) and dot-dashed ($\mu$ model) lines. The points correspond to the velocity dispersion at different mass bins. Both descriptions of the degree of energy equipartition describe the measured dispersion well. However, models using the $\mu$ parameter capture the variation of sigma  at the edges of the mass range better than those using $\eta$. 

The “no IMBH/BHS” cluster shows a higher degree of energy equipartition, characterised by a steeper slope than the “BHS” cluster. Both the $\eta$ and $\mu$ parameters have a higher value in the case of the first cluster. The cluster with an IMBH is characterised by a trend opposite to that found in the other systems evolving towards energy equipartition. This system is characterised by the velocity dispersion increasing with stellar mass, corresponding to negative values for $\eta$ and $\mu$ and it is our motivation to introduce the negative range for $\eta$ and $\mu$ summarized in Eq. \ref{eq:new_range}.

\section{Degree of energy equipartition in the simulated GCs}
\label{sec:results}

For each simulated GC in our sample, we estimate the degree of energy equipartition through the best-fit values of $\eta$ and $\mu$ using the discrete likelihood function in Eq \ref{eq:likelihood}; for the full radial extension of our data and for specific regions of the clusters. We separate the GCs sample into four groups depending on the central objects that each cluster contains. The “no IMBH/BHS” corresponds to GCs that do not host an IMBH nor a significant fraction of stellar-mass black holes (less than ten black holes at $12\,\text{Gyr}$). The “BHS” group corresponds to GCs with more than ten stellar-mass black holes at $12\,\text{Gyr}$ ($N_{\text{bh}}>10$); all of the clusters in this group follow the fallback prescription for supernovae natal kicks \citep[see][]{askar_2017}. The “low-mass IMBH” corresponds to GCs that host an IMBH with a mass fraction of $M_{\text{IMBH}}/M_{\text{GC}}\leq2\%$ (hereafter $f_{\text{IMBH}}$). Finally, the “high-mass IMBH” group includes GCs hosting an IMBH with a mass fraction of $f_{\text{IMBH}}>2\%$. Table \ref{tab:gc_groups} summarises how many clusters are in each category.

\begin{table}
    \centering
    \caption{GCs groups based on their central dark component. For each group, we include their representative symbol for figures, the number of simulations included and their selection criteria. We use the number of stellar-mass black holes $N_{\text{bh}}$ and the mass fraction of the IMBH $f_{\text{IMBH}} = M_{\text{IMBH}}/M_{\text{GC}}$ to separate the samples.}
    \begin{tabular}{l|c|c|l}
        \hline
        Group & Symbol & Number of GCs & Criteria \\
        \hline
        no IMBH/BHS & \includegraphics[scale=0.3]{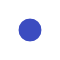} & 34 & \\
        BHS         & \includegraphics[scale=0.3]{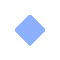} & 26 & $N_{\text{bh}}$ > 10\\
        low-mass IMBH   & \includegraphics[scale=0.3]{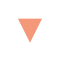} & 14 & $f_{\text{IMBH}}\leq 2\%$\\
        high-mass IMBH  & \includegraphics[scale=0.3]{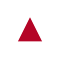} & 27 & $f_{\text{IMBH}}> 2\%$\\
        \hline
    \end{tabular}
    \label{tab:gc_groups}
\end{table}

\subsection{Radial Profiles of the degree of energy equipartition}
\label{sec:radial_profiles}

\begin{figure*}
    \centering
    \includegraphics[width=0.8\linewidth]{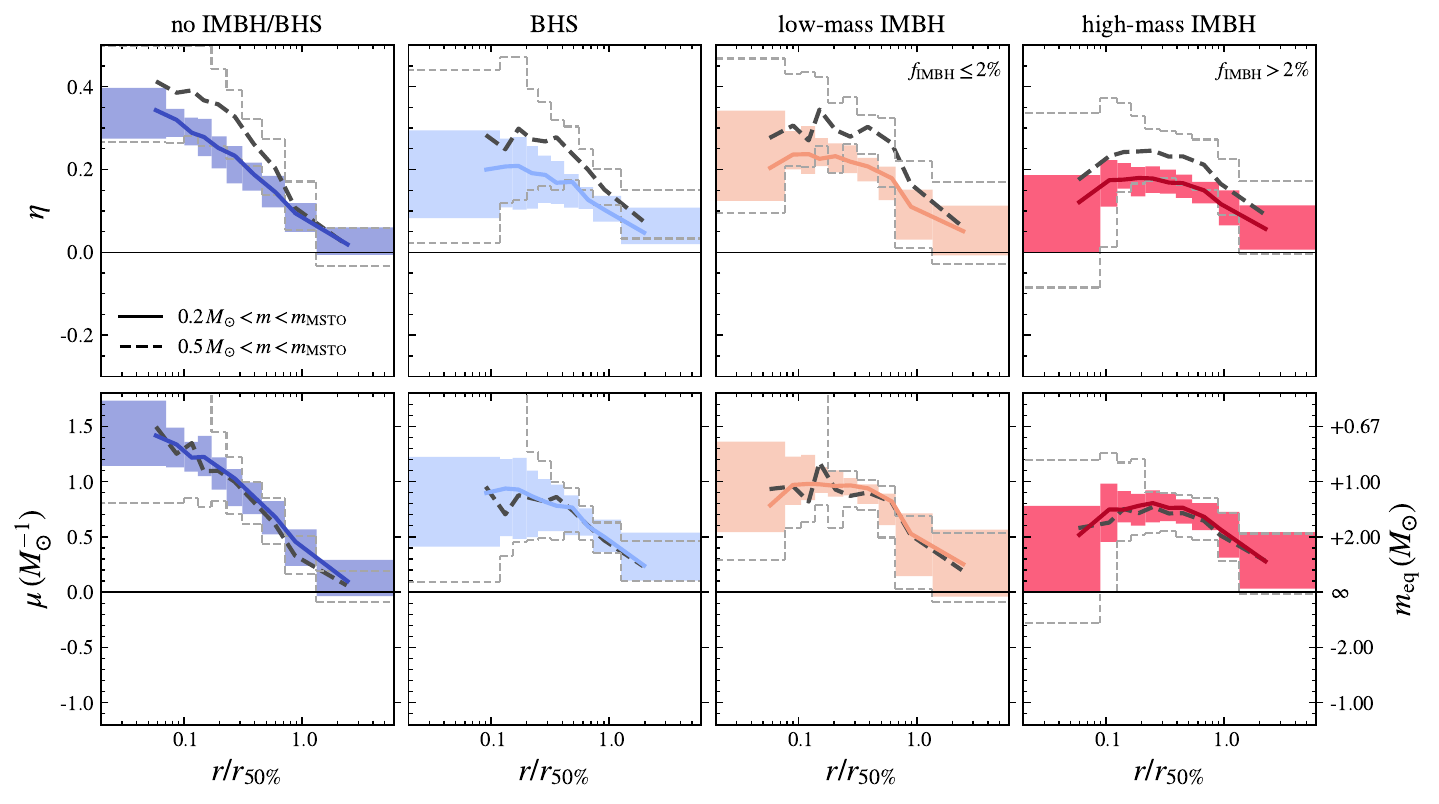}
    \caption{Spherical radial profiles of the degree of equipartition ($\eta$ and $\mu$). Each column represents one of the GC groups introduced in Table \protect\ref{tab:gc_groups}, defined by the presence and types of black holes. The solid lines and coloured shaded regions represent the median and central $80\%$ percentile of the profile distribution for all clusters when using main sequence stars with masses between $0.2\,M_{\odot}$ and the turn-off mass. The dashed lines represent the case for a mass range between $0.5\,M_{\odot}$ and the turn-off mass. For the “no IMBH/BHS” case, both $\eta$ and $\mu$ increase towards the centre as the central regions have a higher degree of energy equipartition. In the case of the “BHS” sample, the median of the distribution shows a lower degree of energy equipartition. Furthermore, the figure shows a higher spread on the $80\%$ percentile than the “no IMBH/BHS” case, representing a significant variation between the clusters. The “low-mass IMBH” case shows similar behaviour to the “BHS” case, but as for the “high-mass IMBH” case, it starts to show a shift to lower values of $\eta$ and $\mu$ in the innermost bin, rather than increasing or flattening profiles. For all groups, we also show that models following the $\eta$ parameterisation are more susceptible to the mass range than models using the $\mu$ parameterisation. We include the corresponding equipartition mass ($m_{\text{eq}}$) values for the range of $\mu$ on the right side of the bottom row. Radial distances are normalized to the 3D half-mass radius, $r_{50\%}$.}     
    \label{fig:radial_profiles_3d}
\end{figure*}

Figure \ref{fig:radial_profiles_3d} shows the $\eta$ and $\mu$ radial profiles for all four groups of GCs, considering spherical 3D coordinates. In each group, the solid and dashed lines represent the median value of all GCs in the group, while the shaded area shows the central $80\%$ percentile range of the combined distribution. In these profiles, we used the masses and velocities of each star in the cluster directly from the $12\,\text{Gyr}$ snapshot to characterise the intrinsic degree of energy equipartition (i.e without adding any kinematic error). For all clusters, radial distances were normalised to their half-mass radius ($r_{50\%}$). We analysed two samples for each GC, one considering all main sequence stars with masses between $0.2\,M_{\odot}$ and the turn-off mass (solid lines and shaded area), and the second considering stars in the range between $0.5\,M_{\odot}$ and the turn-off mass (dashed lines).

GCs without an IMBH nor a BHS show a steady increase in the degree of energy equipartition towards the centre, with an innermost median value of $\eta = 0.35$ and $\mu = 1.42\,M_{\odot}^{-1}$ ($m_{\text{eq}}=0.7\,M_{\odot}$), for the whole mass range. We take the behaviour of the “no IMBH/BHS” sample as the result of the expected evolution of a GC and use it as the fiducial case. As the cluster evolves, its centre achieves a higher degree of energy equipartition, since the inner regions have shorter local values of the relaxation timescale. We obtain a higher value of $\eta$ at each given radius for the profiles using only the bright end of the main sequence (masses between $0.5\,M_{\odot}$ and the turn-off mass). The difference between the mass samples results from the model representation of the degree of equipartition. A broken power-law better captures the mass dependency of the velocity dispersion than a single power-law. The introduction by \cite{bianchini_2016} of the equipartition mass ($m_{\text{eq}}$) and a different functional form capturing broken power law shape provides a more consistent result for both mass samples. In the bottom panel left panel of Figure \ref{fig:radial_profiles_3d}, the estimations of $\mu$, which follows the equipartition mass parameterisation, show how both median profiles are consistent for the “no IMBH/BHS” sample. The systematic differences found when using the $\eta$ model for different mass ranges are present in all four groups.

\subsection{Differences between the models}
\label{sec:diff_rad_prf}
The radial profiles shown in Figure \ref{fig:radial_profiles_3d} show some significant differences between models in the different groups we have identified and clearly illustrate the dynamical effects of stellar BHs and IMBH in the evolution towards energy equipartition. Specifically, the degree of energy equipartition reached by our models in the central regions is shown to depend on the presence of BHs and IMBH in the cluster's core.

GCs in the “BHS” sample have a lower central degree of equipartition than for the “no IMBH/BHS” case. The profiles distribution median has a central value of $\eta=0.2$ and $\mu=0.9\,M_{\odot}^{-1}$ ($m_{\text{eq}}=1.1\,M_{\odot}$). In contrast with the “no IMBH/BHS” case, the “BHS” cluster shows a spread on the central $80\%$ of the profiles distribution, which indicates a clear variation in the degree of energy equipartition between the clusters in the sample. This spread in the values of $\eta$ and $\mu$ is related to the stellar mass black hole content on each GC, which we discuss further in Section \ref{sec:num_bh}.

We find a distinct behaviour for the samples with a central IMBH. While a couple of them are consistent with the “BHS” cases, the majority show a turn-over profile on which the degree of equipartition decreases towards the cluster centre and in particular within the $1\%$ Lagrangian radius (calculated without including the mass of the IMBH). The “low-mass IMBH” group includes a combination of models, some consistent with a turn-over profile and others with a flat/increasing profile. The spread in the innermost bin for the $80\%$ profile distribution shows this diversity of energy equipartition profiles. For clusters in the “high-mass IMBH” group, the turn-over profiles dominate the sample, and most GCs show a lower central value of $\eta$ and $\mu$. 

\begin{figure*}
    \centering
    \includegraphics[width=0.8\linewidth]{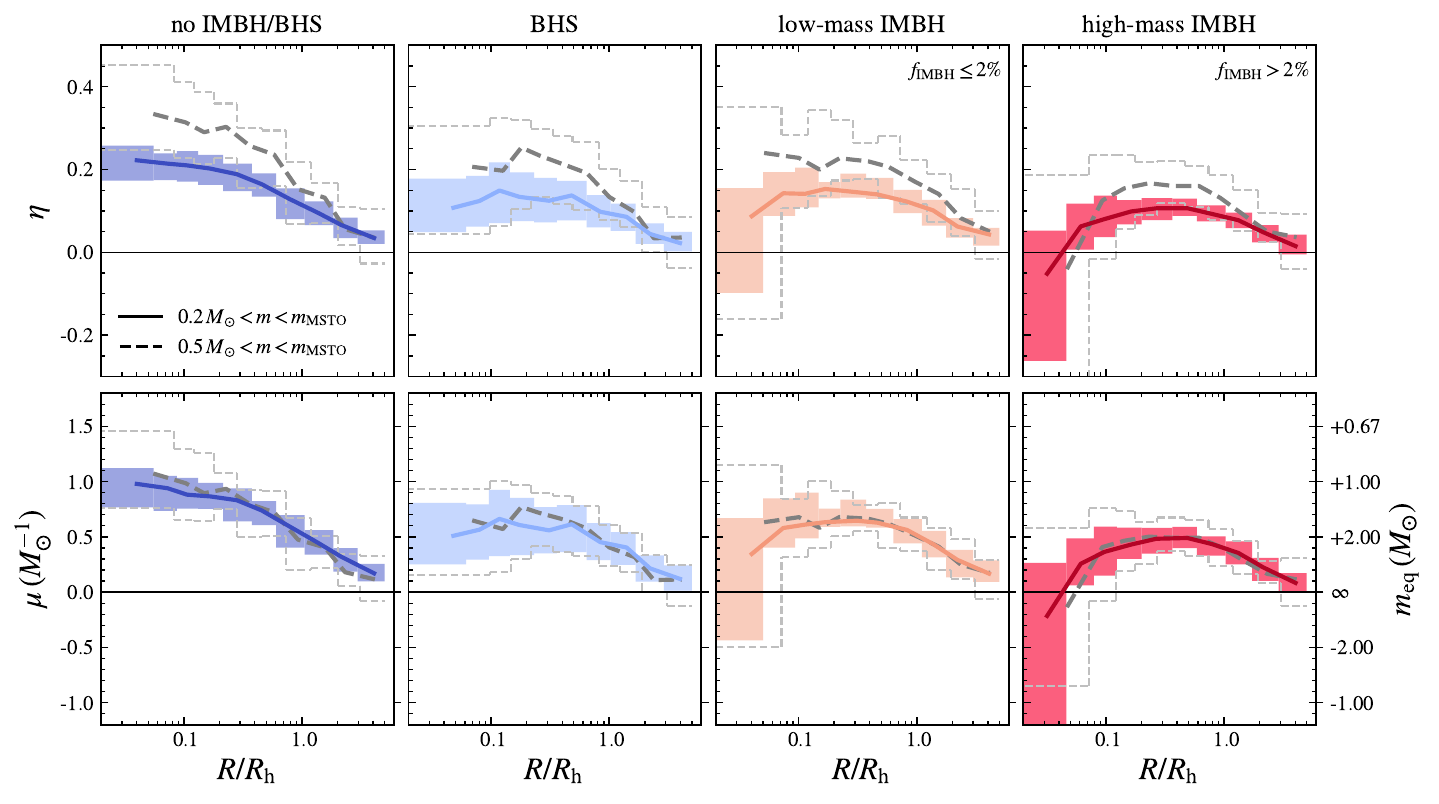}
    \caption{Projected radial profiles of the degree of energy equipartition ($\eta$ and $\mu$). As in Figure \protect\ref{fig:radial_profiles_3d}, we show the distribution of projected radial profiles for $\eta$ and $\mu$. The projected profiles show lower $\eta$ and $\mu$ values than the spherical case (Figure \protect\ref{fig:radial_profiles_3d}), given that we include stars in the observed line-of-sight column with different spherical radii. For both groups with an IMBH, the turnover in the central regions is more significant and arrives at negative values of $\eta$ and $\mu$, as in the case of panel (c) in Figure \protect\ref{fig:example_fit}.  In the projected case, we also see that models based in $\eta$ are susceptible to the mass range. Radial distances are normalized to the 2D half-light radius, $R_h$}
    \label{fig:radial_profiles_2d}
\end{figure*}

We continue our analysis by focusing our attention on the radial variation of $\eta$ and $\mu$ calculated using 2D projected radial distances and proper motion velocities. This allows us to explore the extent to which the effect revealed by our 3D analysis in Figure \ref{fig:radial_profiles_3d} may be detected in observational studies. To reduce stochastic effects, the profile of each cluster is calculated as the median of the profiles obtained by sampling multiple line-of-sights. Figure \ref{fig:radial_profiles_2d} shows that the projected radial profiles of $\eta$ and $\mu$ follow the same trends found in the 3D profiles and reveal similar differences among the four groups in which we have divided our models. As expected, the maximum values of the $\eta$ and $\mu$ parameters are smaller than those found in the 3D analysis; this is the effect of projection as the line-of-sight column includes stars from different spherical radii for a single projected radius. This effect is more significant near the cluster centre than its outer regions since in the outer regions the projected radius samples more closely the spherical radius. Even with the projection effects on the measurements of the degree of energy equipartition, the stronger central equipartition of the “no IMBH/BHS” clusters compared to that of the other models with a BHS and an IMBH is still visible.

It is interesting to point out that GCs with an IMBH may be characterized by a negative value of $\eta$ and $\mu$ in the innermost radial bin; this is the case in particular for models with an IMBH with a mass fraction larger than $2\%$. From the spherical profiles, we know that the intrinsic degree of energy equipartition can be lower at the innermost radial bin but stay positive. In this case, the measured negative values of $\eta$ and $\mu$ result from projection effects. GCs with a high-mass IMBH have a characteristic velocity dispersion cusp in their centre due to the keplerian potential from the IMBH. Stars in the cluster centre will have a much larger velocity dispersion than stars in the outer parts of the cluster, and while partly suppressed \citep[see][]{gill_2008}, some mass segregation is still present in our sample of simulated GCs hosting an IMBH. The combination of these two effects keeps a high velocity dispersion for the high-mass stars while decreasing the velocity dispersion of the low-mass stars as they preferentially populate the outer regions of the cluster. The lower velocity dispersion of outer low-mass stars with projected distance close to the cluster centre leads to an apparent trend of velocity dispersion increasing with mass resulting in the negative values of $\eta$ and $\mu$ shown in panel (c) of Figure \ref{fig:example_fit}. 

\begin{figure*}
    \centering
    \includegraphics[width=0.8\linewidth]{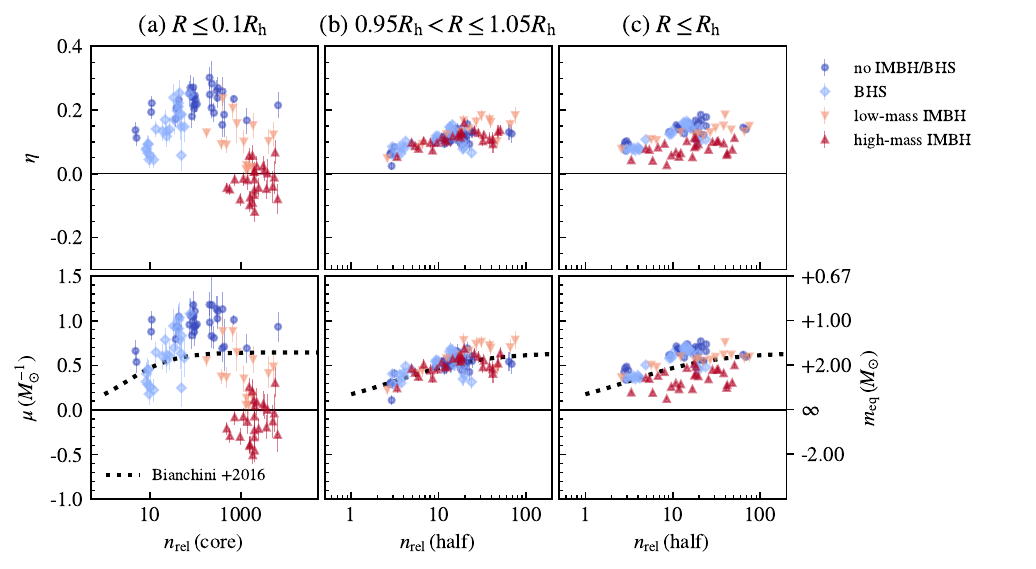}
    \caption{Degree of energy equipartition versus the dynamical age. For each cluster, we show the degree of energy equipartition ($\eta$ and $\mu$) for three regions: the central $10\%$ of the half-light radius, around the half-light radius and all stars within the half-light radius. Once again, we use main-sequence stars within $0.2\,M_{\odot}$ and the turn-off mass. In the first case, we use the current number of relaxation times at $12\,\text{Gyr}$ based on the core relaxation time, while for the other two, we use the half-light relaxation time. In the inner regions, we see two distinctive groups: one including the “no IMBH/BHS” and the “BHS” clusters, and the other with systems in the “high-mass IMBH”. The “low-mass IMBH” sits in between these groups. In the case of the “no IMBH/BHS” and “BHS” clusters, the degree of energy equipartition increases with dynamical age. In contrast, the “high-mass IMBH” group has systematically lower values of $\eta$ and $\mu$ for the same dynamical age, driven by the turnover profile shown in Figure \protect\ref{fig:radial_profiles_2d}. Sampled stars around the half-light radius do not show any clustering and, though there is some scattering, most clusters follow an increasing degree of energy equipartition with dynamical age. Combining all sampled stars within the half-light radius shows the general increasing trend with dynamical age. The “high-mass IMBH” group shows the same behaviour but with an offset at lower values of degree of equipartition. The dashed line in the bottom row represents the equipartition mass and dynamical age relation from \protect\cite{bianchini_2016}.}
    \label{fig:relaxation times}
\end{figure*}

\subsection{Effects of dark remnants and dynamical age on energy equipartition.}
\label{sec:dyn_age}

In order to further explore the effects of dark remnants on the evolution towards energy equipartition and compare the evolution of the different models considered in our study, in this section we  focus on three specific regions:  the innermost region contained within the $10\%$ of the half-light radius, a shell around the half-light radius and all stars within the half-light radius. The first two regions trace the shape of the radial profiles, while the latter will allow us to see if we can observe the effect in the radial profiles when the degree of energy equipartition is measured by including all stars over a large region. We estimate the degree of energy equipartition for each area using the same discrete approach for the projected profiles but with a single line-of-sight projection, allowing us to compare the simulations with Galactic GCs \footnote{The single line-of-sight allows to include the effect of kinematic errors and possible stochastic effects. However, we observe that the latter is not significant. The lowest number of stars in the inner $10\%$ of the half-light radius is $427$ stars, but the measured degree of equipartition is consistent with the multiple line-of-sight radial profile. The range of numbers of stars within $0.1R_{\text{h}}$ ranges from $427$ to $10917$. The majority of clusters in our sample have between $1000$ and $4000$ stars within  $0.1R_{\text{h}}$.}.

We compare the degree of energy equipartition of different clusters as a function of their dynamical age, where we use their current relaxation time as a tracer. To do so, we measure the relaxation time in the core and half-light radius following the expressions in \cite{bianchini_2016}, which are based on \cite{djorgovski_1993} (core relaxation time) and \cite{binney_tremaine_2008} (half-light relaxation time). We measure the core radius as the point where the projected number density is half its central value, for which we first fit a power-law function to the surface number density of each cluster using all the stars in the snapshot \citep[see][]{aros_2020}. For some of the clusters in our sample, particularly those with an IMBH, the projected number density has a cusp that limits us from defining a precise core radius as it depends on how close to the centre we measure the central number density. We have defined the innermost radius so that the central projected number density is given by the its value at $R=0.01R_{\text{h}}$ for all clusters. 

Figure \ref{fig:relaxation times} shows the degree of equipartition as a function of the number of relaxation times ($n_{\text{rel}}$) in the core and half-light radius. We define $n_{\text{rel}}$ as the current snapshot time of $12\,\text{Gyr}$ divided by the respective relaxation time measured at $12\,\text{Gyr}$. While all simulated clusters have the same stellar age, they have different dynamical ages depending on their initial conditions and subsequent evolution. We expect that the degree of energy equipartition increases with dynamical age until a maximum value and then stays constant or decreases as the cluster undergoes core collapse \citep[see ][]{trenti_2013,bianchini_2018,pavlik_2022}. 

In column (a) of Figure \ref{fig:relaxation times}, we see that such is the behaviour of the “no IMBH/BHS” sample, which increases in $\eta$ and $\mu$ as a function of $n_{\text{rel}}(\text{core})$ with a maximum value of $\eta$ and $\mu$ around $n_{\text{rel}}(\text{core})\sim100$. A couple of clusters with $n_{\text{rel}}(\text{core})>100$ show a lower value of $\eta$ and $\mu$. GCs hosting a BHS populate the region consistent with lower dynamical ages and degree of equipartition, extending to the left of the “no IMBH/BHS” sample. GCs hosting a BHS are dynamically younger, given that stellar-mass black holes act as an energy source for the cluster \citep[see e.g.][]{wang_2016,kremer_2018,kremer_2020,bhat_2023}. However, the region populated by the “BHS” sample is not exclusive, and we see a couple of “no IMBH/BHS” GCs that are also dynamically young. These clusters are at the massive end of the initial conditions range with $1.2\times10^6$ stars and about $M_{t=0}\sim7\times10^5\,M_{\odot}$.
Their initial density profiles follow those of \cite{king_1966} models with central dimensional potential $W_0=3$ and $W_{0} =6$, and initial half-mass relaxation times of $\sim 3\,\text{Gyr}$. We also find that the increasing value of $\eta$ and $\mu$ for the “BHS” with dynamical age depends on the number of hosted stellar-mass black holes (see discussion in Section \ref{sec:num_bh}). GCs hosting an IMBH move away from the expected dynamical evolution to a lower degree of equipartition at large $n_{\text{rel}}(\text{core})$, with the “high-mass IMBH” having the most extreme behaviour and the “low-mass IMBH” sample sitting in between the expected dynamical evolution trend and the high-mass IMBHs. 

Column (b) of Figure \ref{fig:relaxation times} shows the degree of equipartition at the half-light radius as a function of the number of half-light relaxation times ($n_{\text{rel}}(\text{half})$). All clusters follow an increasing degree of equipartition with dynamical age. There is no significant distinction for the different cases except for clusters hosting and IMBH with $n_{\text{rel}}(\text{half})>25$, where the “low-mass IMBH” and  “high-mass IMBH” samples split. Column (c) shows that several clusters in the “high-mass IMBH” sample keep part of their central degree of equipartition signature, and are located parallel to the “no IMBH/BHS” and “BHS” clusters at a lower value of degree of equipartition. 

For all cases in Figure \ref{fig:relaxation times}, we include in the panels showing the values of $\mu$ a comparison with the equipartition mass as a function of the dynamical age relation found by \citeauthor{bianchini_2016} (\citeyear{bianchini_2016}; see that paper for further detail about this relation).

\section{Dynamical effects of massive objects}
\label{sec:dyn_bh}

In section \ref{sec:results} we showed that the presence of stellar or intermediate-mass black holes in GCs impacts their evolution towards  energy equipartition. Here we further explore these effects and show how the number of stellar-mass black holes and the properties of the IMBH impact the measured values of $\eta$ and $\mu$. 

\subsection{Predicting the velocity dispersion of stellar-mass black holes.}
\label{sec:bhs_velocity}

We discussed in Section \ref{sec:radial_profiles}, particularly for Figure \ref{fig:radial_profiles_3d}, that models based on the $\eta$ parameter are sensitive to the mass range used in the estimation. A single power law fails to describe the shape of the mass-dependent velocity dispersion, which motivated \cite{bianchini_2016} to introduce the $m_{\text{eq}}$ parameterisation (Eq. \ref{eq:meq_original}), which adds a mass dependence in the log-slope of the mass-dependent velocity dispersion. 

\begin{figure}
    \centering
    \includegraphics[width=0.8\linewidth]{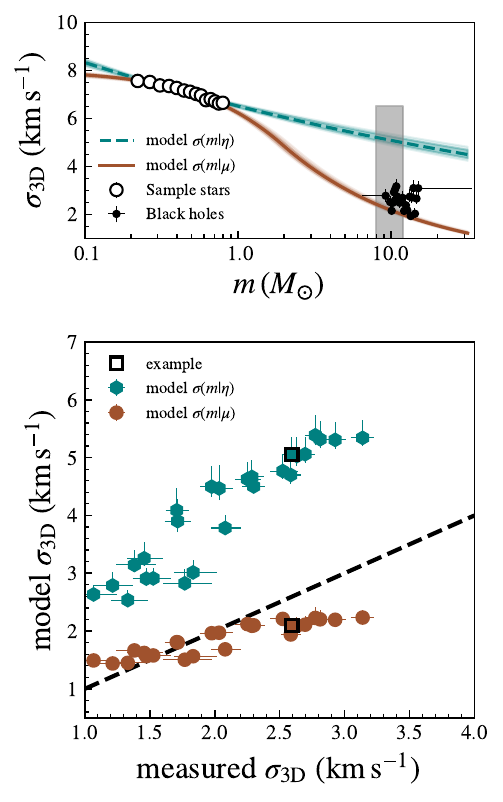}
    \caption{\textit{Top panel}: Example case for the mass-dependant velocity dispersion for a GC with $460$ stellar-mass black holes within the central $5\%$ Lagrangian radius (tridimensional). The panel shows the best-fit models for the degree of energy equipartition ($\eta$ and $\mu$, dashed and solid lines, respectively) given by main-sequence stars within $0.2\,M_{\odot}$ and the MSTO mass (white circles). For comparison, we include the velocity dispersion of the population of stellar-mass black holes in the clusters. The model parameterization using $\mu$ provides a better prediction for the black holes velocity dispersion than the $\eta$ parameterization. \textit{Bottom panel}: Comparison of the measured velocity dispersion of the black hole population (between $8-12\,M_{\odot}$) and the predicted velocity dispersion from the $\eta$ (hexagons) and $\mu$ (circles) parameterizations, for the tridimensional case. The dashed line shows the 1-1 relation, while the black square represents the example in the top panel. For all clusters hosting stellar-mass black holes, the $\mu$ parameterisation provides a better prediction for the black hole's velocity dispersion than the $\eta$ parameterisation, which systematically overestimates it.}
    \label{fig:model_bhs}
\end{figure}

In Figure \ref{fig:model_bhs} we compare the predicted velocity dispersion of the black hole population given by both parameterisations of the degree of energy equipartition, when only using single main-sequence stars to estimate the values of $\eta$ and $\mu$ (with masses between $0.2\,M_{\odot}$ and $m_{\text{MSTO}}$). We focused on stars and black holes within the $5\%$ lagrangian radius using the tridimensional coordinates without kinematic errors (as in Figure \ref{fig:radial_profiles_3d}). The top panel shows an example of a GC with $\sim460$ stellar-mass black holes (from which $\sim420$ are within the analysed radius). While both parameterisations are a good fit for the tracer stars, we observe a significant difference in velocity dispersion outside that mass range. For masses in the range of the stellar-mass black holes, the difference is about $\sim3\,\text{km}\,\text{s}^{-1}$ in velocity dispersion. 

To quantify this difference, we compare the predicted velocity dispersion for both model parameterisations with the measured velocity dispersion of the black hole population for all clusters in the “BHS” sample. For each cluster, we take only the mass range of $8-12\,M_{\odot}$ of stellar-mass black holes as this is the mass range for the first BHs in the GC; more massive black holes are the product of collisions, mergers and binary interactions \citep[see][]{askar_2017}. We use the velocity dispersions at $m=10\,M_{\odot}$ from the best-fit model of $\eta$ and $\mu$ using the tracer stars. The bottom panel of Figure \ref{fig:model_bhs} shows the predicted velocity dispersion from the models versus the measured velocity dispersion from the stellar-mass black holes. While neither model can accurately predict the velocity dispersion of the stellar-mass black holes, the offset on the $\eta$ model is larger than for the $\mu$ parameterisation, with a mean offset of $\sim100\%$ for the $\eta$ models and $\sim14\%$ for the $\mu$ models. We also observe a correlation with the number of black holes in the samples. For clusters with $N_{\text{bh}}\sim20$ the $\mu$ model overestimates the velocity dispersion, while for clusters with $N_{\text{bh}} > 200$, the $\mu$ model underestimates the measured velocity dispersion. 

\begin{figure}
    \centering
    \includegraphics[width=0.8\linewidth]{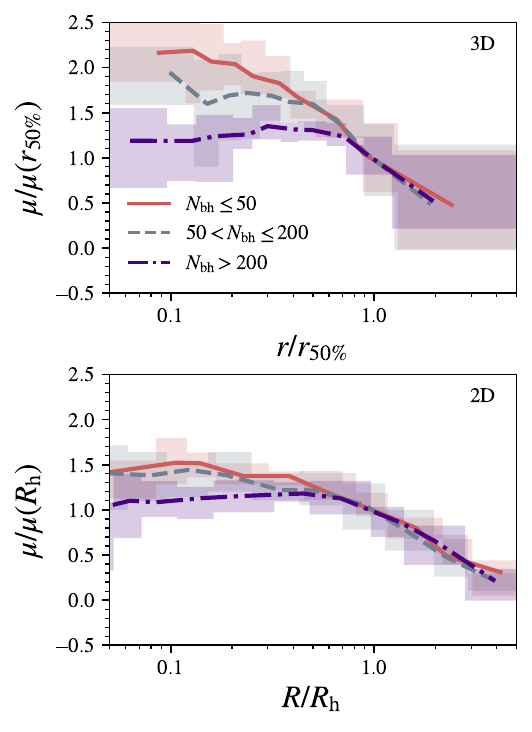}
    \caption{Tridimensional (top) and projected (bottom) radial profiles for the degree of energy equipartition traced by the $\mu$ parameterisation for three subsamples with a different number of stellar-mass black holes. The profiles in both panels were normalised to the value of $\mu$ at the half-mass and half-light radius, respectively. The tridimensional profiles show a different behaviour of the $\mu$ profile for each subsample, particularly for those clusters with a high number of stellar-mass black holes, with a flatter profile within the half-light radius. These differences are less apparent in the projected case, and the dependence on the number of stellar-mass black holes is milder than the tridimensional case.}
    \label{fig:num_bhs}
\end{figure}

\subsection{Dependence of the degree of energy equipartition on the number of stellar-mass black holes}
\label{sec:num_bh}

In the discussion of Figure \ref{fig:radial_profiles_3d}, we indicate that the width of the profile distribution is a consequence of the number of stellar mass-black holes in the clusters. We took three subsamples of the radial profiles for the degree of equipartition given the number of black holes in them: $N_{\text{bh}}\leq50$, $50<N_{\text{bh}}\leq200$ and $N_{\text{bh}}>200$. Figure \ref{fig:num_bhs} shows the three subsamples for the tridimensional and projected case, normalised by their value at the 3D half-mass radius, $r_{50\%}$, and half-light radius, $R_{\text{h}}$, respectively. For both projections, the relative value of $\mu$ in the centre decreases with the number of stellar-mass black holes, which is also observed in Figure \ref{fig:mass_fraction_imbh} when comparing the values of $\mu$ and $\eta$ at the cluster projected centre with the cluster mass fraction in black holes. Within the half-mass radius, the profiles for the subsample with more than $200$ stellar-mass black holes are mostly flat. All GCs in this subsample have retained between $400$ and $500$ black holes, accounting for about $1\%$ of the cluster mass. 

\subsection{Effects of IMBH's properties in the degree of energy equipartition.}
\label{sec:IMBH}

For clusters hosting an IMBH we focus on two properties of the black holes: the mass fraction and the time of formation of the IMBH ($f_{\text{IMBH}}$ and $t_{\text{IMBH}}$ respectively). The first traces how dominant the IMBH is in the central potential, while the second shows the co-evolution impact of the IMBH in the GC. Note that the formation and growth of an IMBH follow after the simulated clusters achieve a central mass density larger than $10^6\,M_{\odot}\,\text{pc}^{-3}$ and is triggered either at an early phase of the GC evolution for initially dense culsters, or later during the core-collapse phase as long the GC have managed to keep a single stellar-mass black hole seed \citep[see][]{giersz_2015}. In our sample, the simulation with the latest IMBH formation is one in which the IMBH forms at $t_{\text{IMBH}}=11.15\,\text{Gyr}$, about $850\,\text{Myr}$ before the observed snapshot. At $12\,\text{Gyr}$ the cluster's core relaxation time is $t_{\text{r,core}}=2.3\,\text{Myr}$ (and $n_{\text{rel}}(\text{core})=289.4$) and the IMBH has a mass of $M_{\text{IMBH}}=1607.7\,M{\odot}$ and mass fraction of $f_{\text{IMBH}}=0.88\%$. This cluster has recently undergone core-collapse, which triggered the IMBH formation.

\begin{figure}
    \centering
    \includegraphics[width=0.8\linewidth]{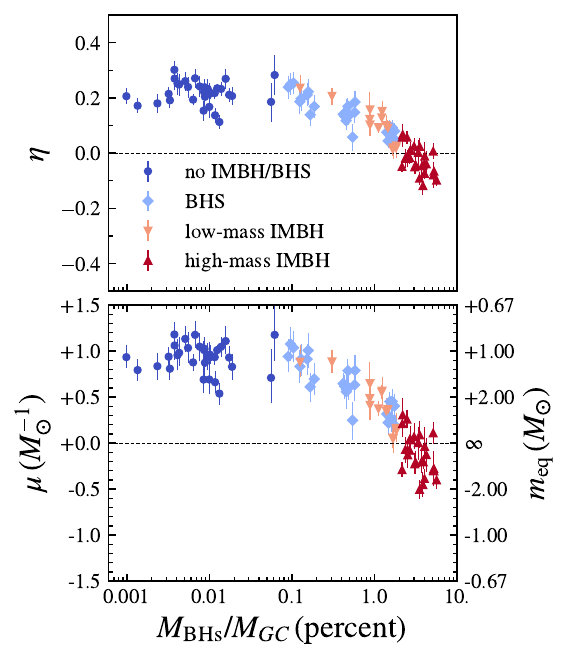}
    \caption{Comparison of the degree of energy equipartition within $10\%$ of the half-light radius with the mass fraction in black holes (stellar- or intermediate-mass). The degree of energy equipartition decreases with increasing mass fraction for both parameterisations, with the “high-mass IMBH” having the largest impact. The “BHS” and “low-mass IMBH” samples overlap in both mass fraction and degree of energy equipartition.}
    \label{fig:mass_fraction_imbh}
\end{figure}

Figure \ref{fig:mass_fraction_imbh} shows the projected $\eta$ and $\mu$ values, for stars within $10\%$ of the half-light radius and masses between $0.2\,M_{\odot}$ and $m_{\text{MSTO}}$, as a function of the mass fraction of the total mass in black holes (either stellar or intermediate-mass black holes). For values of the mass fraction larger than $0.1\%$, the degree of energy equipartition decreases for increasing mass fractions. We notice that the “low-mass IMBH” sample overlaps with the “BHS” sample; however, the “low-mass IMBH” sample have a significantly smaller core and are dynamically older than clusters in the “BHS” sample, as shown in Figure \ref{fig:relaxation times}. The smaller core and lower core relaxation time can help to distinguish between clusters with stellar-mass black holes and low-mass IMBHs. 

\begin{figure}
    \centering
    \includegraphics[width=0.8\linewidth]{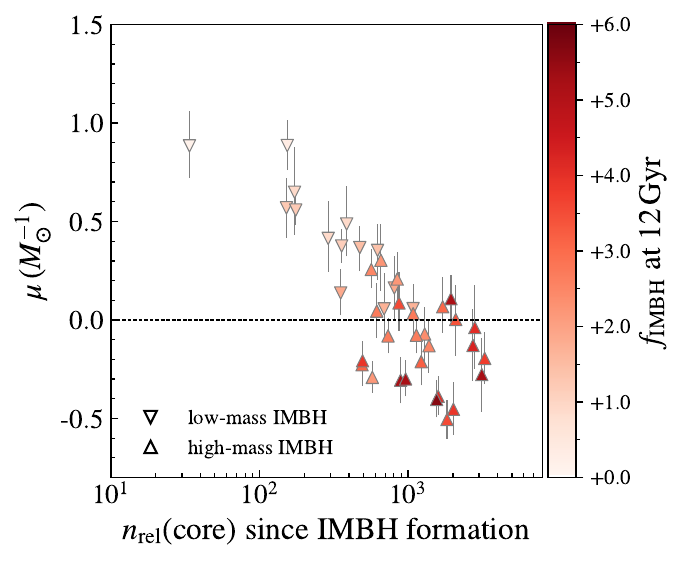}
    \caption{Projected degree of energy equipartition $\mu$ within $0.1\,R_{\text{h}}$ as a function of the number of core relaxation times since IMBH formation. The two samples of IMBHs are represented by downwards-pointing triangles (“low-mass IMBH”) and upper-pointing triangles (“high-mass IMBH”). The markers are colour coded by the IMBH mass fraction. The impact of the IMBH on the degree of energy equipartition depends on how long the IMBH co-evolve with its surrounding stars, which is also tightly connected with the mass growth of the IMBH. The formation time is defined as when the IMBH seed black hole passes $100\,M_{\odot}$, and the older IMBHs had a long time to accrete mass and have larger mass fractions.}
    \label{fig:imbh_t0_v2}
\end{figure}

To explore how the co-evolution time of the IMBH and its host GC affects the degree of energy equipartition in the cluster, we calculate the number of core relaxation times passed from the time of formation of the IMBH (i.e. the time when the seed black hole has a mass of $100\,M_{\odot}$) until the observed snapshot at $12\,\text{Gyr}$. Figure \ref{fig:imbh_t0_v2} shows the 2D degree of energy equipartition $\mu$ within $0.1\,R_{\text{h}}$, as a function of the number of core relaxation times since the IMBH formation. After the formation of the IMBH, it takes time for the new IMBH to grow, dominate its surroundings and change the degree of energy equipartition. We see in Figure \ref{fig:imbh_t0_v2} that the observed degree of energy equipartition decreases for older IMBHs. This is the case for the low-mass IMBHs, while the high-mass IMBHs have a similar behaviour but with a larger scatter. As in Figure \ref{fig:mass_fraction_imbh}, the IMBHs with larger mass fractions have the most substantial impact on $\mu$ and also are the ones that have co-evolved the longest with their host GC.

\begin{figure}
    \centering
    \includegraphics[width=0.9\linewidth]{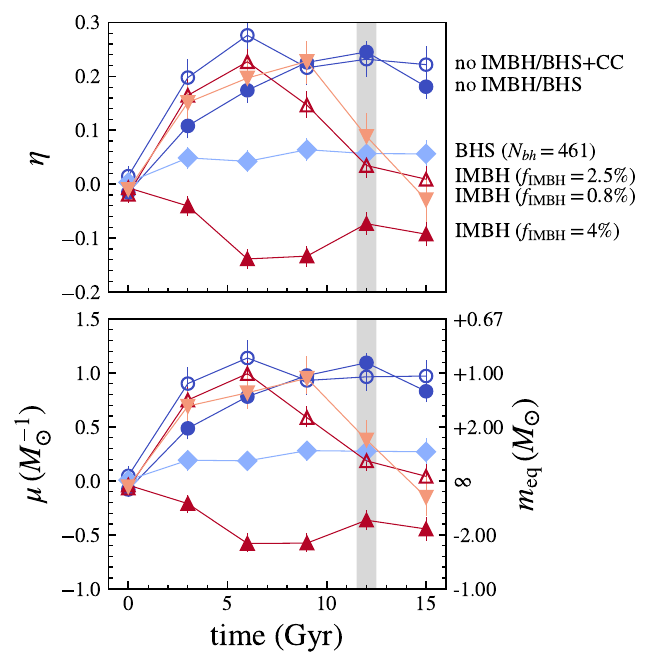}
    \caption{Time evolution of the projected degree of energy equipartition ($\eta$ and $\mu$) in the central $10\%$ of the half-light radius for six GCs in our sample, characterising the different subsamples. For each cluster, we labelled the type of central objects in the top panel. We have included an additional simulation without primordial binaries that achieve core collapse within the measured time (“no IMBH/BHS+CC”). We added a vertical grey region to highlight the $12\,\text{Gyr}$ snapshot used for all Figures. All clusters start without any degree of energy equipartition at time zero. The early-formed high-mass IMBH dominates the central cluster dynamics, and we measure a negative value of $\eta$ and $\mu$ at all times. The “BHS” cluster stays at the same degree of equipartition after the appearance of the stellar-mass black holes. Both clusters without BHS nor an IMBH show a growing degree of energy equipartition with time. The two late IMBHs initially follow the “no IMBH/BHS” clusters until the formation of the central IMBH, where the degree of energy equipartition starts to decrease. }
    \label{fig:time_evolution}
\end{figure}

We selected six simulations to exemplify the time evolution of the degree of equipartition given the central objects in the GC, particularly to show how long it takes, after the late formation of an IMBH, to change the degree of energy equipartition. Figure \ref{fig:time_evolution} shows the values of $\eta$ and $\mu$ in the cluster centre for the projected case at different times\footnote{For all clusters and at each epoch we follow the same approach as for the $12\,\text{Gyr}$ case: we take single stars within $0.2\,M_{\odot}$ and $0.9\,M_{\odot}$ within $10\%$ of the current projected half-light radius. The fixed mass range helps to compare the different epochs as the $m_{\text{MSTO}}$ changes overtime due to the stellar ages.}. We highlighted the current snapshot at $12\,\text{Gyr}$ as a grey vertical band.

We included an additional simulation without primordial binaries, which reaches  core-collapse earlier than any of our “no IMBH/BHS” clusters (labelled as “no IMBH/BHS+CC” ). The two cases with neither an IMBH nor a BHS have a growing degree of energy equipartition with time. The “no IMBH/BHS+CC” case has a slightly more rapid evolution towards energy equipartition probably due to the more rapid evolution towards a denser central structure. The “BHS” cluster is characterized by an early mild increase in the degree of energy equipartition which then remains constant over time as all stellar-mass black holes retained in the cluster form within the first $100\,\text{Myr}$ and halt the cluster's subsequent evolution towards energy equipartition. The early “high-mass IMBH” has a mass fraction of $4\%$ at $12\,\text{Gyr}$ and was formed at  $t_{\text{IMBH}}=11.4\,\text{Myr}$. This IMBH dominated the central dynamics of the cluster at all times, and we measured a negative value of $\eta$ and $\mu$ at each epoch. Finally, we have two cases of late formation: one within the “low-mass IMBH” ($t_{\text{IMBH}}=11.1\,\text{Gyr}$) sample and the other in the “high-mass IMBH” ($t_{\text{IMBH}}=7.5\,\text{Gyr}$) sample. Both clusters follow the same growth in the central degree of energy equipartition as the “no IMBH/BHS” GC until the formation of the IMBH. Once the IMBH starts growing in mass and begins to dominate the central dynamics, the values of $\eta$ and $\mu$ decrease in the centre. As these IMBHs form during the cluster’s core collapse, they might appear as a core-collapsed cluster but have a lower degree of energy equipartition than a cluster that does not host an IMBH and has undergone core-collapse. 

\section{Comparison with Galactic globular clusters}
\label{sec:obs}

\begin{figure}
    \centering
    \includegraphics[width=0.8\linewidth]{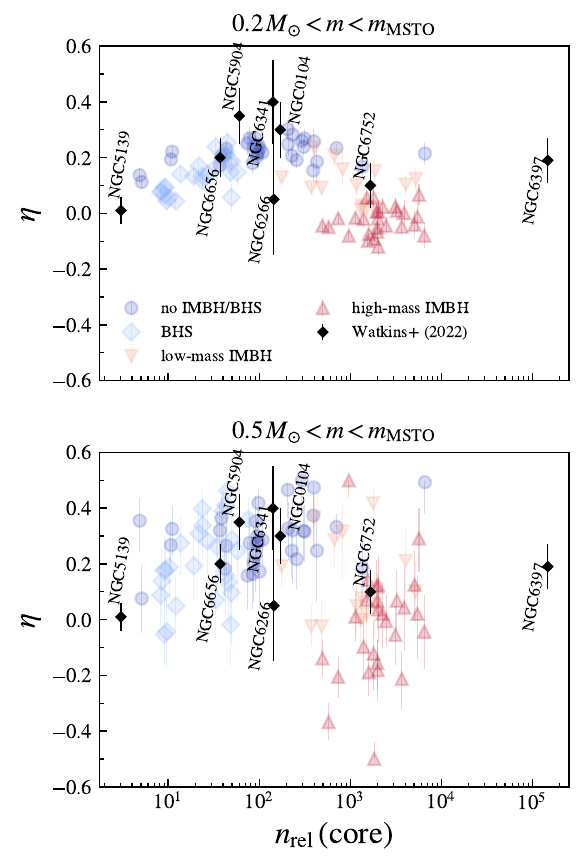}
    \caption{Comparison of the degree of energy equipartition as a function of the dynamical age in simulations with galactic GCs. We show the measured $\eta$ values for eight Galactic GCs from \protect\cite{watkins_2022}, taken from their degree of energy equipartition radial profiles. We compare them with the $\eta$ values from the simulated GC sample considering the $(0.2\,M_{\odot},m_{\text{MSTO}})$ mass range (as in the top-left panel of Figure \protect\ref{fig:relaxation times}) and the $(0.5\,M_{\odot},m_{\text{MSTO}})$ mass range.}
    \label{fig:nrelax_comparison_observations}
\end{figure}

Although the models analysed in this work are not aimed at specifically fitting any particular Galactic globular cluster \citep[see e.g.][for specific models for NGC 104 and NGC 5139]{giersz_2003,giersz_2011}, it is interesting to compare the central values of $\eta$ for a sample of eigth Galactic globular clusters from the radial profiles in Figure 14 of \cite{watkins_2022} with our results. For these clusters, we use the number of core relaxation times presented in Table 1 in \cite{watkins_2022}. Figure \ref{fig:nrelax_comparison_observations} shows the comparison of the Galactic GCs with the $\eta$ values for our full mass range (top panel) and with stars with masses between $0.5\,M_{\odot}$ and the MSTO mass (bottom panel). We include the second sample as the mass range for the Galactic GCs varies between clusters. The GCs NGC 104, NGC 5904 and NGC 6341 are consistent with the “no IMBH/BHS” samples, particularly for the bottom panel (NGC 5904 and NGC 6341 have mass ranges between $\sim 0.4\,M_{\odot}$ and $\sim 0.9\,M_{\odot}$). NGC 6397, on the other hand, is dynamically older than any of the simulated GCs in our samples, and it is a post core-collapse GC. A more extended sample of simulated GCs might be necessary to find comparable cases.

NGC 6656 is consistent with dynamically older models in the “BHS” sample with $N_{\text{bh}}\sim20-30$ stellar-mass black holes at $12\,\text{Gyr}$. These simulated clusters are also the ones with the highest degree of energy equipartition in the BHS sample (see Figure \ref{fig:num_bhs} and discussion therein). \cite{strader_2012} found two stellar-mass black holes in NGC 6656, and different studies have suggested that the total number of stellar-mass black holes in the clusters is between $16$ and $57$ \citep[see e.g.][]{sippel_2013,heggie_2014,askar_2018,weatherford_2020}. On the other hand, NGC 5139 is consistent with the dynamically younger side of the “BHS” sample (small $n_{\text{rel}}(\text{core})$). While the origin of the central dark mass of NGC 5139 is still an open problem, recent works suggest the presence of multiple stellar-mass black holes in its core \citep{zocchi_2019,baumgardt_2019}. The region populated by the “BHS” sample is not exclusive and a couple of models in the “no IMBH/BHS” sample also populate the same region. These models are clusters dynamically younger due to their initial high mass and shallow density profiles. From this comparison, we cannot exclude the possibility of NGC 5139 simply being a massive and dynamically young cluster \cite[see also][]{giersz_2003}. We included in Figure \ref{fig:nrelax_comparison_observations} the value of $\eta$ for the innermost bin of NGC 5139 from Figure 14 of \cite{watkins_2022}. Outside the innermost bin, the degree of energy equipartition is consistent with a constant value of $\eta\sim0.1$. If we take this value instead for the central bin, our previous statements do not change, and NGC 5139 is still consistent with both the “BHS” sample and the dynamically young “no IMBH/BHS” clusters.  

NGC 6266 and NGC 6752 are in the region between the “no IMBH/BHS” and “high-mass IMBH”, and are consistent with some of the “low-mass IMBH”. Previous studies have estimated the mass of a possible IMBH in NGC 6266, finding upper limits of $M_{\text{IMBH}}<4\times 10^3\,M_{\odot}$ \citep{mcnamara_2012}, a value of $M_{\text{IMBH}}=2\pm1\times 10^3\,M_{\odot}$ \citep{lutzgendorf_2013a} or no IMBH at all \citep{baumgardt_2017}. From our comparison with the simulations, we cannot exclude either possibility; while it appears to be consistent with the “low-mass IMBH” sample, the large error bars in $\eta$ also makes it compatible with the “no IMBH/BHS” sample. NGC 6752 is in a similar situation, while previous studies have suggested the presence of a binary system hosting a $M_{\text{IMBH}}\sim 200\,M_{\odot}$ IMBH \citep{ferraro_2003,colpi_2003}, recent dynamical analysis shows that no IMBH is necessary \citep{baumgardt_2017}. NGC 6752 is a post core-collapse \citep{djorgovski_1993} and none of our “no IMBH/BHS” clusters has undergone the deep core-collapse typical of systems with no (or very small fraction) of primordial binaries. Further dynamical analysis of this cluster is necessary (Scalco et al. in prep.).

\section{Summary}
\label{sec:summary}

We analysed $101$ simulated GCs from the \textsc{MOCCA-Survey} Database I \citep{belloni_2016,askar_2017} to explore the effect of stellar-mass and intermediate-mass black holes in the degree of energy equipartition. The simulated GCs have different initial conditions and can have a central intermediate-mass black hole (IMBH), a stellar-mass black hole system (BHS) or neither at $12\,\text{Gyr}$. We applied two commonly used parameterisation models to measure the degree of energy equipartition \citep[][]{trenti_2013,bianchini_2016} of each cluster’s stellar sample to analyse its radial behaviour and the dynamical signatures of the presence of an IMBH or BHS.

We show (see Figures \ref{fig:radial_profiles_3d} and \ref{fig:radial_profiles_2d}) that the shape of the radial variation of the degree of energy equipartition depends on the kind of central objects in the cluster. Clusters that do not have multiple stellar-mass black holes nor a central IMBH (“no IMBH/BHS”) have an increasing degree of energy equipartition towards the cluster centre. A higher degree of energy equipartition at the centre is expected since the local relaxation time decreases at smaller clustercentric distances. The measured profiles are also consistent with previous studies \citep[see][]{trenti_2013}. 

Clusters with stellar-mass black holes (labelled as “BHS” and with $N_{\text{bh}} > 10$) have a systematically lower degree of energy equipartition in the cluster centre than those in the “no IMBH/BHS” sample. We find that the number of stellar-mass black holes (and their total mass fraction relative to the cluster mass) also affects the degree of energy equipartition, where the degree of energy equipartition is lower for increasing number of black holes in the cluster (see Figure \ref{fig:num_bhs}).
%where the effect increases with the number of black holes in the cluster (see Figure \ref{fig:num_bhs}).
While GCs in the “BHS” sample are, in general, dynamically younger than those in the “no IMBH/BHS”, we find examples of clusters in the “no IMBH/BHS” sample that have a lower value in the degree of energy equipartition and are dynamically younger than the rest of the sample. The dynamically young  “no IMBH/BHS” clusters overlap with those in the “BHS” group and show that these dynamical signatures are not exclusive of the presence of many stellar-mass black holes (see Figure \ref{fig:relaxation times}).

We find that the presence of the IMBH can reduce the degree of energy equipartition in the cluster centre and produce a turn-over profile in which the central values are lower than at the half-mass or half-light radius (see Figures \ref{fig:radial_profiles_3d} and \ref{fig:radial_profiles_2d}). This “turn-over” feature is more significant with increasing mass-fraction of the central IMBH, particularly for the projected profiles where the degree of energy equipartition can have negative values at the cluster centre (see Figure \ref{fig:mass_fraction_imbh}). The lower values for the degree of energy equipartition and the smaller cores separate clusters with an IMBH from the expected evolutionary path of the cluster’s central regions (see Figure \ref{fig:relaxation times}). We also find that the signature in the degree of energy equipartition is more apparent the longer the IMBH co-evolves with its host GC (see Figure \ref{fig:imbh_t0_v2}). 

Our work shows that IMBHs and stellar-mass black holes reduce the central degree of energy equipartition in GCs. Previous works have shown that IMBH and stellar black holes act as an energy source in the cluster but disentangling the combined effects of black holes on the cluster structure,  mass segregation, and  energy exchanges between black holes and main sequence stars resulting in the reduced degree of energy equipartition is a more complex task; this requires further investigation following in detail the dynamics of encounters and energy exchanges in the presence of black holes in the clusters central regions.

As a side product of our analysis, we show that the parameterisation based on the equipartition mass \citep[][]{bianchini_2016} is more robust than using a single power law when using different mass ranges (see Figures \ref{fig:radial_profiles_3d} and \ref{fig:radial_profiles_2d}). Furthermore, the models based on the equipartition mass can better predict the expected velocity dispersion of the stellar-mass black hole population (see Figure \ref{fig:model_bhs}). 

We compared our results with a sample of Galactic GCs from \cite{watkins_2022} and found that our results from the simulations are consistent with the Galactic GCs. Two clusters, NGC 6266 and NGC 6752 appear to be consistent with the anomalous behaviour of simulated GCs with an IMBH. On the other hand, NGC 5139 and NGC 6656 are consistent with the sample of clusters hosting multiple stellar-mass black holes. However, we find that the properties of NGC 5139 are also consistent with models of  massive and dynamically young clusters that do not have many stellar-mass black holes ($N_{\text{bh}}\sim 2-3$ at $12\,\text{Gyr}$). We cannot exclude the possibility of NGC 5139 being dynamically young without many stellar-mass black holes.  

During our analysis, we considered only single-main sequence stars and assumed that the mass of each star is known. In the case of an observational sample, an isochrone fitting is necessary for estimating the mass of each star \citep{watkins_2022}. The isochrone fitting method works well for single stars, but binary stars might have their masses underestimated. By assigning the wrong mass, the low-velocity tail of the velocity distribution at a given mass is populated and a bias to the corresponding velocity dispersion is introduced. We explored this issue by measuring the degree of energy equipartition at the cluster centre with binaries and using their real and estimated masses. While our colour selection eliminates most binaries, the ones that remain in the sample can still add a bias to the measurement of the degree of energy equipartition. We find that the samples with the wrong mass assignment are biased towards a higher degree of energy equipartition with increasing binary fraction. However, the differences in the degree of energy equipartition between the two samples are around $10\%$ to $20\%$ in the worst cases, while their estimation errors are still consistent with no difference at all. Selecting stars close to isochrone reduces the binary contamination significantly; the primary source of the remaining contamination can come from a binary system with a main-sequence star and a white dwarf. 

For our calculation of the degree of energy equipartition we have used the total velocity dispersion and have not studied the possible difference between the degree of energy equipartition in the different velocity components \citep[see][for a study exploring the development of these differences particularly at large clustercentric distances]{pavlik_2021}. While the analysis of equipartition in the different velocity components is beyond the scope of this paper, we will further explore this issue in a future investigation.

The analysis presented in this paper provides the theoretical framework necessary to link the degree of energy equipartition to the evolutionary history and current dynamical state of GCs. Our results can guide the interpretation of current and upcoming observational studies and provide the foundation for future investigations in which we will further explore the evolution towards energy equipartition in systems with broader initial conditions, including anisotropy and the presence of multiple stellar populations.

%The last numbered section should briefly summarise what has been done, and describe
%the final conclusions which the authors draw from their work.

\section*{Acknowledgements}

EV and FIA acknowledge support from NSF grant AST-2009193. We thank the referee for their constructive comments and suggestions.  FIA thanks Abbas Askar and the MOCCA team for making the simulation library available, and also thanks Vaclav Pavlik, Alexander Livernois and Michele Scalco for helpful discussions. 
This research was supported in part by Lilly Endowment, Inc., through its support for the Indiana University Pervasive Technology Institute. This research also made use of the \texttt{NUMPY} package \citep{van_der_walt_2011}, \texttt{EMCEE} \citep{emcee_2013} package and \texttt{ASTROPY} \citep{astropy_2022} package, while all figures were made using \texttt{MATPLOTLIB} \citep{hunter_2007}.

%%%%%%%%%%%%%%%%%%%%%%%%%%%%%%%%%%%%%%%%%%%%%%%%%%
\section*{Data Availability}

The data for the simulated GCs used in this article was provided by the MOCCA\footnote{\url{https://moccacode.net/}} group and it will be shared upon request with permission of the MOCCA group. 

%The inclusion of a Data Availability Statement is a requirement for articles published in MNRAS. Data Availability Statements provide a standardised format for readers to understand the availability of data underlying the research results described in the article. The statement may refer to original data generated in the course of the study or to third-party data analysed in the article. The statement should describe and provide means of access, where possible, by linking to the data or providing the required accession numbers for the relevant databases or DOIs.



\bibliographystyle{mnras}
\bibliography{energy-eq} % if your bibtex file is called example.bib

%\begin{thebibliography}{99}
%\end{thebibliography}

%%%%%%%%%%%%%%%%%%%%%%%%%%%%%%%%%%%%%%%%%%%%%%%%%%

%%%%%%%%%%%%%%%%% APPENDICES %%%%%%%%%%%%%%%%%%%%%

\appendix
\section{Kinematic Errors}
\label{app:errors}

\begin{figure*}
    \centering
    \includegraphics[width=0.8\linewidth]{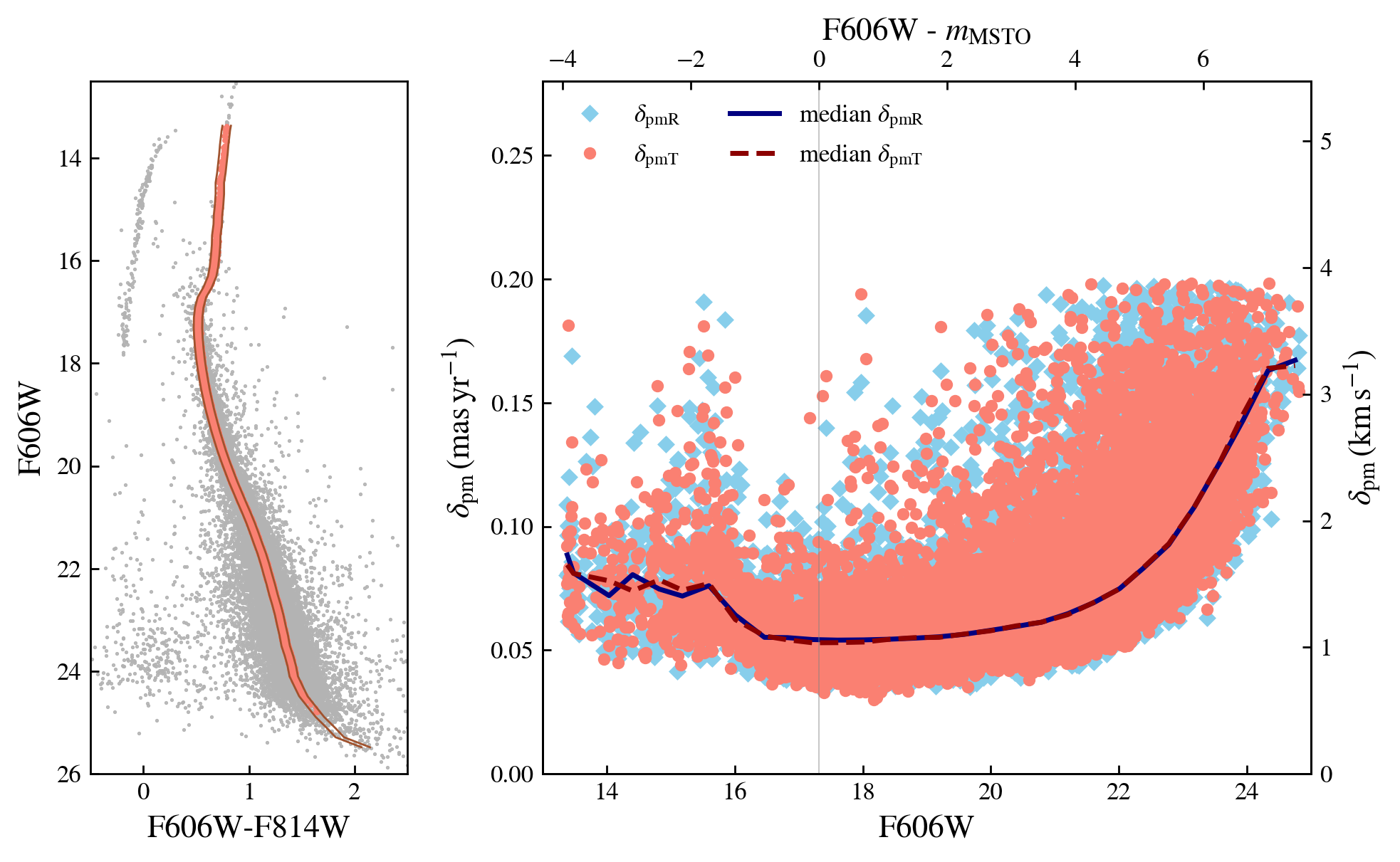}
    \caption{Colour-Magnitude diagram and proper motion errors for NGC 6752. The left panel show the colour-magnitude diagram for all the stars in NGC 6752 from the HACKS catalogue. The dark-red lines show the colour range given by 0.02mag from the median colour, and the light red points show our selection of stars to characterise the median velocity errors. The right panel shows the radial (light-blue squares) and tangential (light-red circles) proper motion errors in both $\text{mas}\,\text{yr}^{-1}$ and $\text{km}\,\text{s}^{-1}$ (right y-axis at the distance of NGC 6752). We use the median errors (blue-solid and red-dashed lines) to assign kinematic errors to the simulations.}
    \label{fig:kin-err}
\end{figure*}

We add kinematic errors to the simulation data following the error distribution from HST data for the cluster NGC 6752; we chose this cluster as its magnitude range covers a similar region to the one given for the stellar mass limits we analyse in this work ($0.2\,M_{\odot}$ to $\sim 0.9\,M_{\odot}$ which is equivalent to the range within the MTSO and seven magnitudes below the MSTO). The kinematic and photometric data for NGC 6752 comes from the HACKS database \citep[][available at \url{https://archive.stsci.edu/hlsp/hacks}]{libralato_2022}. We made a selection in proper motions and colours to sample possible member stars within the main sequence of the cluster. To do so, we select all stars with velocities within $\sim4\sigma$ and those within a colour range of $0.02\,\text{mag}$ from the median colour (F606W-F814W). The left side panel of Figure \ref{fig:kin-err} shows all the stars in the HACKS catalogue of NGC 6752 in grey and the selected stars in light-red.
 
We transformed the RA and DEC proper motions and their observational errors to radial and tangential proper motions ($v_{\text{pmR}}$ and $v_{\text{pmT}}$) to find their median error value at a given magnitude bin. The side right panel of Figure \ref{fig:kin-err} shows the proper motion errors as a function of the F606W magnitudes and the median values (blue-solid and red-dashed lines). With the median error profiles, we assign a mock kinematic error and an observational random noise to each star in the simulation sample.

The final “observed” proper motions for the stars in the simulated GCs are:
\begin{align}
v_{\text{pmR},\text{obs}} =& v_{\text{pmR},\text{sim}} + N(0,\delta_{\text{pmR}}(\text{mag}-\text{mag}_{\text{MSTO}}))\\
v_{\text{pmT},\text{obs}} =& v_{\text{pmT},\text{sim}} + N(0,\delta_{\text{pmT}}(\text{mag}-\text{mag}_{\text{MSTO}}))\,,
\end{align}   
where $N(0,\delta_{\text{pmR}})$ and $N(0,\delta_{\text{pmT}})$ are the added observational random noise, sampled from a normal distribution with zero mean and dispersion equal to the kinematic error of the star. This noise is important as the likelihood function (Eq. \ref{eq:likelihood}) assumes that the intrinsic velocity dispersion and the observational errors contribute to the total dispersion in the observed velocities.

%%%%%%%%%%%%%%%%%%%%%%%%%%%%%%%%%%%%%%%%%%%%%%%%%%

% Don't change these lines
\bsp	% typesetting comment
\label{lastpage}
\end{document}